\documentclass[twocolumn,a4paper,reprint,aps,prd,amssymb,superscriptaddress,nofootinbib]{revtex4-1}
\usepackage{graphicx,xcolor}
\usepackage{epsf}
\usepackage{bm}
\usepackage{amsmath}
\usepackage{amsfonts}
\usepackage{amssymb}
\usepackage{epstopdf}
\usepackage{natbib}
\usepackage{nameref}
\usepackage{color}
\usepackage{verbatim}
\usepackage{multirow}
\usepackage{bm}
\usepackage{dcolumn}
\usepackage{placeins}

\usepackage{subfigure}
\definecolor{darkblue}{rgb}{0.0, 0.0, 0.55}
\definecolor{darkred}{rgb}{0.55, 0.0, 0.0}
\usepackage{hyperref}
\hypersetup{
    colorlinks=true, 
    linkcolor=darkblue,
    citecolor=darkblue,
    urlcolor=darkblue}
    
\makeatletter\let\expandableinput\@@input\makeatother

\begin{document}

\title{Probing late-time deviations from \texorpdfstring{$\Lambda$CDM}{LCDM} with a quadratic dark energy expansion}

\author{Sehjal Khandelwal}
\email{sehjal.khandelwal@plaksha.edu.in}
\affiliation{Plaksha University, Mohali, Punjab-140306, India}

\author{Abraão J. S. Capistrano}
\email{capistrano@ufpr.br}
\affiliation{Universidade Federal do Paran\'{a}, Departmento de Engenharia e Exatas, Rua Pioneiro, 2153, Palotina, 85950-000, Paraná/PR, Brasil\\
Federal University of Latin American Integration (UNILA), Applied physics graduation program, Avenida Tarqu\'{i}nio Joslin dos Santos, 1000-Polo Universit\'{a}rio, Foz do Igua\c{c}u, 85867-670, Paran\'{a}/PR, Brasil}

\author{Suresh Kumar}
\email{suresh.kumar@plaksha.edu.in}
\affiliation{Plaksha University, Mohali, Punjab-140306, India}

\author{Rafael C. Nunes}
\email{rafadcnunes@gmail.com}
\affiliation{Instituto de F\'{i}sica, Universidade Federal do Rio Grande do Sul, 91501-970 Porto Alegre RS, Brasil}
\affiliation{Divisão de Astrofísica, Instituto Nacional de Pesquisas Espaciais, Avenida dos Astronautas 1758, São José dos Campos, 12227-010, São Paulo, Brasil}
\begin{abstract}
We investigate the observational viability of a quadratic dark energy expansion (QDEE) model as a phenomenological extension of the standard $\Lambda\mathrm{CDM}$ cosmological framework. This approach introduces the additional degrees of freedom that permit mild late-time deviations from a constant dark-energy component while preserving the standard early-Universe behavior. We constrain the model using a comprehensive combination of cosmological datasets, including Planck 2018 cosmic microwave background (CMB) measurements, Atacama Cosmology Telescope (ACT) Data Release 6 (DR6) and South Pole Telescope (SPT-3G) data, Dark Energy Spectroscopic Instrument (DESI) Data Release 2 (DR2), and the Pantheon Plus type Ia supernova compilation. Our results show that the QDEE framework shifts the inferred Hubble constant toward higher values relative to $\Lambda\mathrm{CDM}$, partially alleviating the tension with local measurements while remaining consistent with early-Universe constraints. Bayesian model comparison indicates strong evidence in favor of this framework over standard $\Lambda\mathrm{CDM}$ across multiple dataset combinations. Posterior predictive checks further demonstrate that the model yields predictions consistent with the observed data within statistical uncertainties.
\end{abstract}

\maketitle

\section{Introduction}

\label{sec:intro}
One of the most dominant questions in cosmology has been the nature of the dark components of the Universe. The quest to determine whether cosmic acceleration is driven by a constant vacuum energy density or a time-varying dark energy (DE) component remains a major open problem in modern physics \cite{Riess_1998, Perlmutter_1999, Copeland:2006wr, weinberg, Peebles:2002gy, Alam_2004}. Nevertheless, the concordance cosmological model describes gravity through Einstein's General Relativity while unifying cold dark matter (CDM) and DE within the standard $\Lambda\mathrm{CDM}$ framework \cite{Planck2018, Peebles1984, Peebles:2002gy, Peebles:2024txt}. In this picture, DE is modeled as a positive cosmological constant $\Lambda$, associated with vacuum energy that exerts negative pressure and drives the accelerated expansion of the Universe. This component is characterized by a barotropic equation-of-state (EoS) parameter $w = -1$, consistent with a constant vacuum energy density throughout cosmic evolution (see \cite{Carroll:2000fy, RevModPhys.61.1} for reviews on vacuum energy and the cosmological constant). It remains the simplest and most successful framework for describing the evolution of the Universe, and has been remarkably successful in explaining a wide range of observations.

Despite its outstanding success in explaining a broad range of cosmological observations, the $\Lambda\mathrm{CDM}$ paradigm continues to face several persistent tensions that may hint at physics beyond the standard concordance model (see \cite{DiValentino:2021izs, Perivolaropoulos:2021jda, Kamionkowski:2022pkx} for reviews). Among these, the Hubble constant ($H_0$) tension remains one of the most pressing challenges in contemporary cosmology. The discrepancy between early- and late-Universe measurements of $H_0$ has reached a statistical significance of approximately $5\sigma$~\cite{Riess:2021jrx}, while more recent results from the H0DN collaboration suggest that this inconsistency may have grown to $7.1\sigma$ \cite{H0DN:2025lyy}. If substantiated, such a deviation would be exceedingly difficult to reconcile within the standard $\Lambda\mathrm{CDM}$ framework.

Additional independent challenges have also emerged from recent baryon acoustic oscillation (BAO) measurements conducted by the Dark Energy Spectroscopic Instrument (DESI)~\cite{DESI:2024mwx, DESI:2025zgx}. In its first two Data Releases (DR1 and DR2), the DESI collaboration reported indications that DE may evolve over cosmic time, rather than behaving as a pure cosmological constant with fixed energy density~\cite{DESI:2024mwx, DESI:2025zgx}. These observations therefore provide a complementary and potentially profound challenge to the $\Lambda\mathrm{CDM}$ scenario.

More specifically, the DESI DR2 BAO dataset shows a preference for dynamical DE at the $2.8\sigma$--$4.2\sigma$ level, with the reconstructed EoS suggesting a transition from a phantom-like phase at earlier epochs to a quintessence-like regime at later times. Such behavior constitutes a significant departure from the standard cosmological constant interpretation. Furthermore, numerous independent studies have reinforced these findings, reporting statistically relevant deviations from the $\Lambda\mathrm{CDM}$ framework and proposing new observational consistency tests using DESI DR2 BAO data~\cite{DESI:2025zgx, DESI:2025zpo, Montani:2025qnk, Capozziello:2025qmh, Ozulker:2025ehg, deCruzPerez:2025dni, Yao:2025wlx, Sohail:2025mma, Smith:2025uaq, Lee:2025axp, Efstratiou:2025iqi, Smith:2025icl, Liu:2025bss, Li:2025muv, RoyChoudhury:2025iis, Chen:2025ywv, Fazzari:2025lzd, Gomez-Valent:2025mfl, Wu:2025vfs, Feleppa:2025clx, Mishra:2025goj, Hussain:2025vbo, Gialamas:2025pwv, Cline:2025sbt, Mukherjee:2025ytj, Bayat:2025xfr, Hussain:2025nqy, Cheng:2025lod, Toomey:2025yuy, Paliathanasis:2025kmg, Luciano:2025hjn, Li:2025vuh, Cheng:2025yue, Benevento:2020fev}, as well as through alternative cosmological datasets \cite{Silva:2025twg, Sabogal:2025jbo, Chudaykin:2025lww, Reeves:2025xau, Ishak:2025cay}. Together, these developments increasingly motivate the exploration of extensions or alternatives to the concordance cosmological model.

One of the simplest yet theoretically well-motivated extensions beyond the $\Lambda\mathrm{CDM}$ paradigm is the quadratic dark energy expansion (QDEE) framework \cite{Sahni_2003, Alam:2003sc, Kumar:2025obb}. This model arises from a truncated Taylor expansion of the total cosmic energy density in powers of $(1+z)$, naturally introducing a quadratic correction to the DE sector. Such a parametrization provides a flexible phenomenological description capable of capturing departures from a pure cosmological constant, including transient phantom-like behavior, through the additional expansion coefficients $\Omega_1$ and $\Omega_2$. In this formulation, the conventional spatial-curvature contribution is effectively replaced by a second-order term, viz., $\Omega_2(1+z)^2$, which shares the same redshift dependence as curvature within the Friedmann equation. We clarify that this contribution may effectively mimic mild curvature-like contributions or geometric degeneracy effects in late-time distance observables. A future detailed comparison with non-flat $\Lambda\mathrm{CDM}$ extensions would be valuable for disentangling genuine dark-energy evolution from effective curvature behavior within the present parametrization. Consequently, this quadratic component may encode the effects of nonzero spatial curvature, an evolving exotic DE sector, or some combination of both \cite{Kumar:2025obb}. More importantly, the reconstructed DE EoS within this framework exhibits a distinctly non-monotonic evolution, often featuring transitions between quintessence-like and phantom-like regimes across cosmic history \cite{Kumar:2025obb}. Together, these higher-order contributions generate an effective late-time negative pressure capable of driving accelerated expansion while simultaneously allowing moderate deviations from the constant-vacuum-energy behavior predicted by $\Lambda\mathrm{CDM}$.

Recent analysis of QDEE scenario \cite{Kumar:2025obb} has combined early-Universe information from Planck CMB observations with low-redshift BAO measurements from DESI, obtaining an inferred Hubble constant of approximately $H_0 \simeq 69.74\pm0.77~\mathrm{km/s/Mpc}$. In the present work, however, we significantly extend this analysis by incorporating a broader and more comprehensive combination of cosmological probes. This expanded joint dataset offers enhanced sensitivity to the redshift dependence of the cosmic expansion history, thereby enabling substantially tighter constraints on the leading QDEE. Moreover, this approach facilitates a more robust comparison of the QDEE framework against competing cosmological scenarios, providing a clearer assessment of its viability as an alternative to the concordance model.

In this work, we present cosmological constraints for late-time DE models, including the standard flat $\Lambda\mathrm{CDM}$ scenario with cold dark matter and a cosmological constant, a time-varying DE EoS model based on the Chevallier Polarski Linder (CPL) parametrization \cite{Chevallier2001, Linder2003}, and a truncated Taylor expansion model for QDEE framework \cite{Sahni_2003, Alam:2003sc, Kumar:2025obb}. The QDEE model is tested as an extension of the standard cosmological framework with other low-redshift datasets. It provides a phenomenological expansion in powers of $(1+z)$. This expansion allows for a mild departure of the DE parameter from a constant vacuum energy at late times.

The paper is structured as follows: In Sec. \ref{model}, we briefly discuss the models analyzed in this work. Sec. \ref{data} describes the datasets used and the methodology followed. Sec. \ref{results} presents the results and discussion of the observational constraints on the model parameters, followed by the conclusions and main findings in Sec. \ref{conclusion}.

\section{Late-Time dark energy Parameterizations}
\label{model}
The primary focus of this work is the QDEE model, which parametrizes the DE sector through a quadratic expansion in $(1+z)$ about the future asymptotic epoch. Unlike EoS parametrizations such as the widely used Chevallier–Polarski–Linder (CPL) model, the QDEE framework directly parametrizes the total cosmic energy density as \cite{Kumar:2025obb}:
\begin{equation}
\label{eqn}
    \rho_{\rm tot}(z) = \sum_{i=0}^4 c_i (1+z)^i,
\end{equation}
where $c_i$ are constant coefficients. Here $\rho_0 \equiv c_0$ corresponds to the vacuum energy density, while $\rho_1 \equiv c_1(1+z)$ and $\rho_2 \equiv c_2(1+z)^2$ capture the leading-order deviations from a constant DE component. The higher-order terms $\rho_3 \equiv c_3(1+z)^3$ and $\rho_4 \equiv c_4(1+z)^4$ scale as standard matter and radiation, respectively. This formulation allows the effective DE EoS to be derived from the underlying energy-density evolution rather than being imposed \textit{a priori}. 

As described in \cite{Kumar:2025obb}, the corresponding effective EoS of the effective DE reads 
\begin{equation}
\label{eos}
w_{\rm de}(z) = -1 + \frac{1}{3}
\frac{\Omega_1(1+z) + 2\Omega_2(1+z)^2}
{\Omega_\Lambda + \Omega_1(1+z) + \Omega_2(1+z)^2},
\end{equation}
with $\Omega_\Lambda \equiv c_0/\rho_{\rm crit,0}$, $\Omega_1 \equiv c_1/\rho_{\rm crit,0}$, $\Omega_2 \equiv c_2/\rho_{\rm crit,0}$, where $\rho_{\rm crit, 0}=\frac{3H_0^2}{8\pi G}$ is the present-day critical density, $G$ is Newton's gravitational constant and $H_0$ is the present-day value of the Hubble parameter. In the asymptotic future limit $z \rightarrow -1$, the model naturally approaches a de Sitter state: $\rho_{\rm tot}(z)\rightarrow\rho_0$ and $w_{\rm de}(z)\rightarrow -1$. Setting $\Omega_1=\Omega_2=0$ recovers the standard flat $\Lambda\mathrm{CDM}$ scenario.

To ensure a physically well-defined, non-singular cosmological evolution for all $z\geq-1$, the conditions $\Omega_0>0$, $\Omega_2>0$, and
$\Omega_0-\frac{\Omega_1^2}{4\Omega_2}>0$,
must be satisfied \cite{Kumar:2025obb}. These conditions are necessary and sufficient to guarantee regularity of the effective EoS over the full physical redshift range. In particular, negative values of $\Omega_1$ remain allowed, thereby permitting a broader class of late-time DE evolution while preserving theoretical consistency. Enforcing these physical priors is essential, as failure to do so would allow pathological regions of parameter space to contaminate posterior distributions and Bayesian evidence estimates, potentially biasing cosmological inference.

For comparison, we also consider the well-established CPL parameterization, which serves as a widely used dynamical DE benchmark. The CPL framework is motivated as the lowest-order Taylor expansion of the DE EoS around the present epoch ($a=1$), offering a simple phenomenological description of time-varying DE \cite{Linder:2006xb}:
\begin{equation}
w(a)=w_0+w_a\left(1-a\right),
\end{equation}
where $w_0$ is the present-day value of the DE EoS, and $w_a$ characterizes its redshift evolution toward earlier cosmic times with $a=1/(1+z)$. The CPL model provides an effective description of the redshift range directly probed by current observations ($z\lesssim2$). However, extrapolating it to very high redshift ($z\gg2$, i.e., $a\rightarrow0$) raises theoretical issues: $w_{\rm de}(z)$ approaches $w_0+w_a$, which can be physically problematic, particularly in the phantom regime $w_0+w_a<-1$. Such behavior can be inconsistent with broader classes of physically motivated scalar-field or barotropic DE models \cite{Artola_2026}. Hence, while CPL remains a valuable observational benchmark, its interpretation as a fundamental DE model beyond the late-time Universe should be treated with caution.

Although both the CPL parametrization and the QDEE model serve as phenomenological extensions of $\Lambda\mathrm{CDM}$ designed to capture possible late-time deviations from a cosmological constant, they differ fundamentally in their underlying construction. The CPL framework directly parametrizes the DE EoS $w(a)$ as a first-order expansion around the present epoch, with the DE density subsequently derived from this assumed form. In contrast, the QDEE model directly parametrizes the DE density itself through a polynomial expansion in $(1+z)$. Consequently, the two models can exhibit similar phenomenological behavior over limited observational redshift ranges, but they generally differ in their high-redshift extrapolation, future asymptotic behavior, theoretical priors, and perturbative consistency. They should therefore be regarded as technically distinct frameworks with different physical assumptions and cosmological implications. A comparison of QDEE with CPL is discussed in \cite{Kumar:2025obb}.

\section{Datasets and Methodology}
\label{data}
To analyze the models, we use the CMB dataset: (i) \textbf{Planck}: We use the Planck likelihood \cite{Planck2018} to combine observations of the CMB temperature and polarization anisotropies. In particular, we use the Plik-lite likelihood for temperature and polarization spectra for $l\geq 30$, which contains observations of spectra up to $l_{\rm max} = 2508$ for TT and $l_{\rm max} = 1996$ for TE and EE and takes into account the impacts of marginalizing over Planck interference and nuisance factors. In accordance with the usual Planck analysis, we employ the Commander likelihood for the TT spectrum and the SimAll likelihood for the EE polarization spectrum for low multiples $2\leq l < 29$ \cite{2020}; (ii) \textbf {ACT}: We utilize the ACT DR6 lensing likelihood \cite{ACT:2023kun, ACT:2023dou}, which provides high-precision measurements of the lensing potential, particularly across intermediate multipole range $40 \leq l \leq 763$. We use the public ACT DR6 lensing likelihood, implemented using the act\_dr6\_lenslike package\footnote{\url{ https://github.com/ACTCollaboration/act_dr6_lenslike}} \cite{Madhavacheril_2024, Qu_2026, Qu_2024, ACT:2023ubw}; (iii) \textbf{SPT}: We utilize the SPT3G\_D1\_TnE\_lite\_candl likelihood\footnote{\url{https://github.com/Lbalkenhol/}} \cite{Ge_2025, Qu_2026}, where the probability is based on temperature and E-mode polarization power spectra obtained from the SPT-3G experiment. We use light compressed covariance without CMB lensing, using internal priors and default data selection. This dataset enhances Planck's sensitivity by restricting small-scale CMB anisotropies. For the SPT-3G D1 TT/TE/EE band-powers, we use publicly available data from \cite{SPT-3G:2025bzu}. We refer to the combined CMB data compilation to as \textbf{CMB}.

In addition to the CMB dataset, we use BAO and supernovae datasets, viz.,
we use measurements from the \textbf{DESI DR2}\footnote{\url{4 https://github.com/CobayaSampler/bao_data}} \cite{Karim_2025}, combining galaxy, quasar, and Lyman-$\alpha$ forest tracers over the redshift range $0.295 \le z \le 2.330$. The BAO information is incorporated through the standard distance ratios $D_{\rm M}/r_{\rm d}$, $D_{\rm H}/r_{\rm d}$, and $D_{\rm V}/r_{\rm d}$, all normalised by the sound horizon at the drag epoch $r_{\rm d}$ \cite{DESI:2025zgx}. We refer to it as \textbf{DESI DR2}.
For supernovae constraints, we use the Pantheon Plus compilation of Type Ia supernovae (SNIa). The Pantheon Plus sample contains 1,701 SNIa spanning the redshift range $0.0012<z<2.26$\footnote{\url{https://github.com/PantheonPlusSH0ES/DataRelease}}, as presented in Scolnic et al. (2022) \cite{Scolnic_2022}. We refer to this dataset as \textbf{PP}.\\

The parameter space explored here for the extended $\Lambda\mathrm{CDM}$ scenarios includes the six common standard cosmological parameters: the present-day physical baryon density $\omega_{\rm b} \equiv \Omega_{\rm b} h^2$, the cold dark matter density $\omega_{\rm cdm} \equiv \Omega_{\rm cdm} h^2$, the angular size of the sound horizon at recombination $\theta_{\rm s}$, the amplitude of primordial scalar perturbations $\ln(10^{10}A_{\rm s})$, the scalar spectral index $n_{\rm s}$, and the optical depth to reionization $\tau_{\rm reio}$. Flat priors are assumed for all parameters, with the following ranges: $\omega_{\rm b} \in [0.018,0.024]$, $\omega_{\rm cdm} \in [0.10,0.14]$, $100\theta_{\rm s} \in [1.03,1.05]$, $\ln(10^{10}A_{\rm s}) \in [3.0,3.18]$, $n_{\rm s} \in [0.9,1.1]$, $\tau_{\rm reio} \in [0.04,0.125]$, $w_0 \in [-3.0,1.0]$, $w_a \in [-3.0,2.0]$, $\Omega_1 \in [-1,0.5]$, and $\Omega_2 \in [0,0.1]$.

The choice of prior support for $\Omega_1$ and $\Omega_2$ is physically motivated by the requirement $\Omega_2 > 0$, which is necessary to prevent singular behavior in the effective DE EoS. These prior intervals are intentionally conservative, restricting the analysis to physically meaningful regions of parameter space \cite{Kumar:2025obb}. 

We implement the cosmological models considered in this work within the \texttt{CLASS} Boltzmann solver\footnote{\url{https://github.com/lesgourg/class_public}}
 \cite{2011arXiv1104.2932L}, allowing the linear perturbations as described in \cite{Kumar:2025obb}, and perform Monte Carlo analyses using the \texttt{MontePython} sampler\footnote{\url{https://github.com/brinckmann/montepython_public}}
 \cite{Audren, Brinckmann}. Convergence of the Markov chains is assessed through the Gelman--Rubin criterion \cite{GelmanRubin1992}, adopting the requirement $R-1 \leq 10^{-2}$. Posterior distributions and statistical summaries are obtained using the \texttt{GetDist} package\footnote{\url{https://github.com/cmbant/getdist}}
 \cite{Lewis:2019xzd}. In addition to parameter constraints and $\chi^2$ statistics, we further assess model consistency through Bayesian evidence and posterior predictive checks (PPC).

\begin{table*}[hbt!]
\caption{Constraints at the 68\% confidence level for the QDEE, CPL, and $\Lambda\mathrm{CDM}$ model parameters. The relative chi-squared minimum and relative log-Bayesian evidence values are also reported for each model, relative to the $\Lambda\mathrm{CDM}$ scenario, viz., $\Delta\chi^2_{\rm min}=\chi^2_{\mathrm{min, model}}-\chi^2_{\rm min,\Lambda\mathrm{CDM}}$ and $\Delta \ln \mathcal{Z} = \ln \mathcal{Z}_{\mathrm{model}} - \ln \mathcal{Z}_{\Lambda\mathrm{CDM}}$. Negative values of $\Delta\chi^2_{\rm min}$ and positive values of $\Delta \ln \mathcal{Z}$ indicate a statistical preference over the standard $\Lambda\mathrm{CDM}$ model.}
\label{constraints}
\resizebox{0.65\textwidth}{!}{
\begin{tabular}{|l|c|c|c|}
\hline
\textbf{Data} & \textbf{CMB} & \textbf{CMB+DESI DR2} & \textbf{CMB+DESI DR2+PP} \\
\hline

\textbf{Model} 
& QDEE
& QDEE
& QDEE \\
& CPL
& CPL
& CPL\\
& $\Lambda\mathrm{CDM}$
& $\Lambda\mathrm{CDM}$
& $\Lambda\mathrm{CDM}$ \\
\hline

{\boldmath$10^{2}\omega_{\rm b}$} 
& $2.228\pm0.013$
& $2.234\pm0.012$
& $2.237\pm0.011$ \\
& $2.233\pm0.012$
& $2.230\pm0.011$
& $2.233\pm0.011$ \\
& $2.230^{+0.012}_{-0.011}$
& $2.244\pm0.011$
& $2.241^{+0.011}_{-0.010}$ \\
\hline

{\boldmath$\omega_{\rm cdm}$}
& $0.12077\pm0.00097$
& $0.11978\pm0.00079$
& $0.11912\pm0.00072$ \\
& $0.12027\pm0.00097$
& $0.12064\pm0.00077$
& $0.12005\pm0.00076$ \\
& $0.12063\pm0.00089$
& $0.11835\pm0.00060$
& $0.11856\pm0.00058$ \\
\hline

{\boldmath$100\theta_{\rm s}$}& $1.04178\pm0.00025$
& $1.04193\pm0.00024$
& $1.04200\pm0.00024$ \\
& $1.04184\pm0.00027$
& $1.04184\pm0.00025$
& $1.04191\pm0.00025$ \\
& $1.04179\pm0.00026$
& $1.04206\pm0.00025$
& $1.04205\pm0.00024$ \\
\hline

{\boldmath$\ln(10^{10}A_{\rm s})$}& $3.0433^{+0.0098}_{-0.0089}$
& $3.0447\pm0.0090$
& $3.0488^{+0.0089}_{-0.0100}$ \\
& $3.0420\pm0.0093$
& $3.0457\pm0.0094$
& $3.0470\pm0.0920$ \\
& $3.0460\pm0.0088$
& $3.0515\pm0.0094$
& $3.0504\pm0.0092$ \\
\hline

{\boldmath$n_{\rm s}$}& $0.9640^{+0.0036}_{-0.0041}$
& $0.9658\pm0.0034$
& $0.9670\pm0.0035$ \\
& $0.9647\pm0.0035$
& $0.9636\pm0.0033$
& $0.9647\pm0.0032$ \\
& $0.9638^{+0.0030}_{-0.0034}$
& $0.9686\pm0.0031$
& $0.9680\pm0.0031$ \\
\hline

{\boldmath$\tau_{\rm reio}$}
& $0.0517\pm0.0044$
& $0.0535\pm0.0043$
& $0.0554\pm0.0046$ \\
& $0.0517^{+0.0046}_{-0.0041}$
& $0.0529\pm0.0044$
& $0.0541\pm0.0045$ \\
& $0.0529\pm0.0045$
& $0.0574\pm0.0044$
& $0.0567\pm0.0043$ \\
\hline

{\boldmath$\Omega_1$}
& $-0.310\pm0.140$
& $-0.140^{+0.056}_{-0.041}$
& $-0.061^{+0.035}_{-0.027}$ \\
& --
& --
& -- \\
& [0]
& [0]
& [0] \\
\hline

{\boldmath$\Omega_2$}
& $0.0330^{+0.0180}_{-0.0300}$
& $0.0084^{+0.0020}_{-0.0084}$
& $0.0049^{+0.0016}_{-0.0049}$ \\
& --
& --
& -- \\
& [0]
& [0]
& [0] \\
\hline

{\boldmath$w_0$}& -- 
& -- 
& -- \\
& $-1.520^{+0.550}_{-0.640}$
& $-0.350^{+0.220}_{-0.170}$
& $-0.813\pm0.054$ \\
& $[-1]$
& $[-1]$
& $[-1]$ \\
\hline

{\boldmath$w_a$}& --
& --
& -- \\
& $-1.10^{+1.41}_{-1.60}$
& $-2.02^{+0.47}_{-0.64}$
& $-0.76\pm0.21$ \\
& [0]
& [0]
& [0] \\
\hline\hline

{\boldmath$\Omega_{\rm m}$}
& $0.2801^{+0.0230}_{-0.0280}$
& $0.2909\pm0.0058$
& $0.3010\pm0.0045$ \\
& $0.1830^{+0.0330}_{-0.0780}$
& $0.3590^{+0.0220}_{-0.0190}$
& $0.3132\pm0.0057$ \\
& $0.3192\pm0.0056$
& $0.3050\pm0.0035$
& $0.3063\pm0.0033$ \\
\hline

{\boldmath$H_0~(\mathrm{km/s/Mpc})$}& $71.8^{+3.20}_{-1.90}$
& $70.07\pm0.73$
& $68.72\pm0.51$ \\
& $92.01\pm20.10$
& $63.30^{+1.5}_{-2.0}$& $67.58\pm0.59$ \\
& $67.08\pm0.40$
& $68.10\pm0.27$
& $67.99\pm0.25$ \\
\hline

{\boldmath$S_8$}
& $0.8250^{+0.0110}_{-0.0130}$
& $0.8229\pm0.0080$
& $0.8210\pm0.0078$ \\
& $0.7670^{+0.0430}_{-0.0520}$
& $0.8590\pm0.0130$
& $0.8365^{+0.0089}_{-0.0076}$ \\
& $0.8390\pm0.0095$
& $0.8160\pm0.0069$
& $0.8180\pm0.0069$ \\
\hline

{\boldmath$r_{\rm d}~(\rm Mpc)$}
& $146.98\pm0.24$
& $147.18\pm0.21$
& $147.32\pm0.20$ \\
& $147.07\pm0.23$
& $147.00\pm0.20$
& $147.12\pm0.21$ \\
& $147.00\pm0.22$
& $147.45\pm0.17$
& $147.43\pm0.18$ \\
\hline\hline

{\boldmath$\Delta\chi^2_{\rm min}$}
& $-9.18$
& $-11.04$
& $-4.32$ \\
& $-10.09$
& $-16.86$
& $-17.20$ \\
& 0
& 0
& 0 \\
\hline

{\boldmath$\Delta \ln \mathcal{Z}$}
& $9.36$
& $9.04$
& $5.77$ \\
& $3.35$
& $6.67$
& $3.72$ \\
& 0
& 0
& 0 \\
\hline

\end{tabular}}
\end{table*}

\section{Results and discussions}
\label{results}

\subsection{Cosmological Parameter Constraints}

The observational constraints on the model parameters for all three models, namely $\Lambda\mathrm{CDM}$, CPL, and QDEE, are summarized in Table \ref{constraints}. The results reveal several important physical and statistical features regarding the viability of the QDEE framework relative to both CPL and standard $\Lambda\mathrm{CDM}$. Generally, the data indicate that while all models remain broadly compatible with current observations, the QDEE scenario consistently provides a more flexible and statistically favored description of late-time cosmology, particularly when evaluated through Bayesian model comparison.

Table \ref{constraints} shows that the six common baseline cosmological parameters remain highly stable across all three frameworks. This demonstrates that neither QDEE nor CPL significantly disrupts the well-established physics constrained by the CMB. In particular, the remarkable consistency in $100\theta_{\rm s}$ confirms that all models preserve the angular acoustic scale at recombination, implying that deviations arise almost entirely from late-time expansion effects rather than modifications to pre-recombination dynamics. The stability of these parameters also indicates that the inclusion of ACT and SPT data in the Planck likelihood does not introduce any significant tension with the base Planck CMB cosmology; instead, it provides tighter constraints and enhanced statistical stability. These results suggest that the additional dark-energy degrees of freedom primarily affect the late-time expansion history.

\begin{figure*}[t]
\centering

\subfigure[]{
    \includegraphics[width=0.48\textwidth]{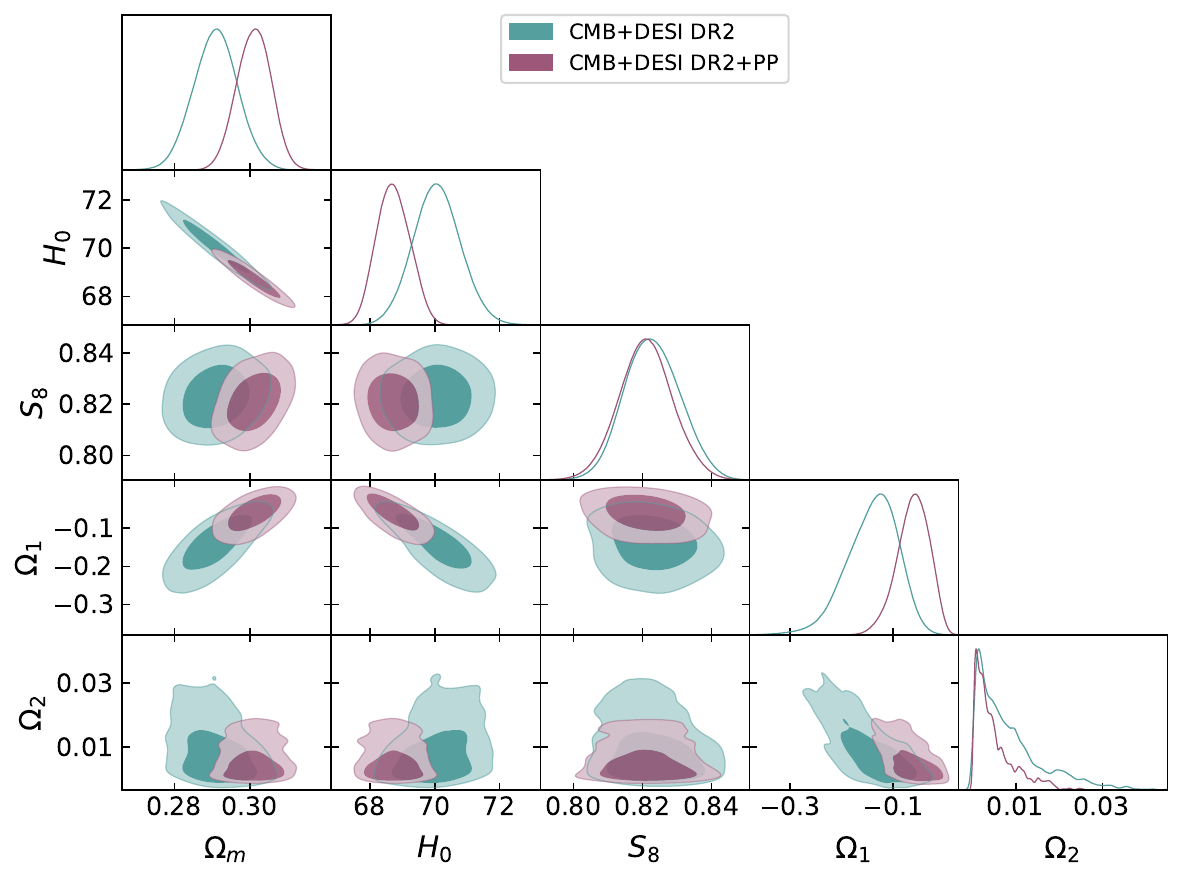}
    \label{fig:taylor_triangle}
}
\hfill
\subfigure[]{
    \includegraphics[width=0.48\textwidth]{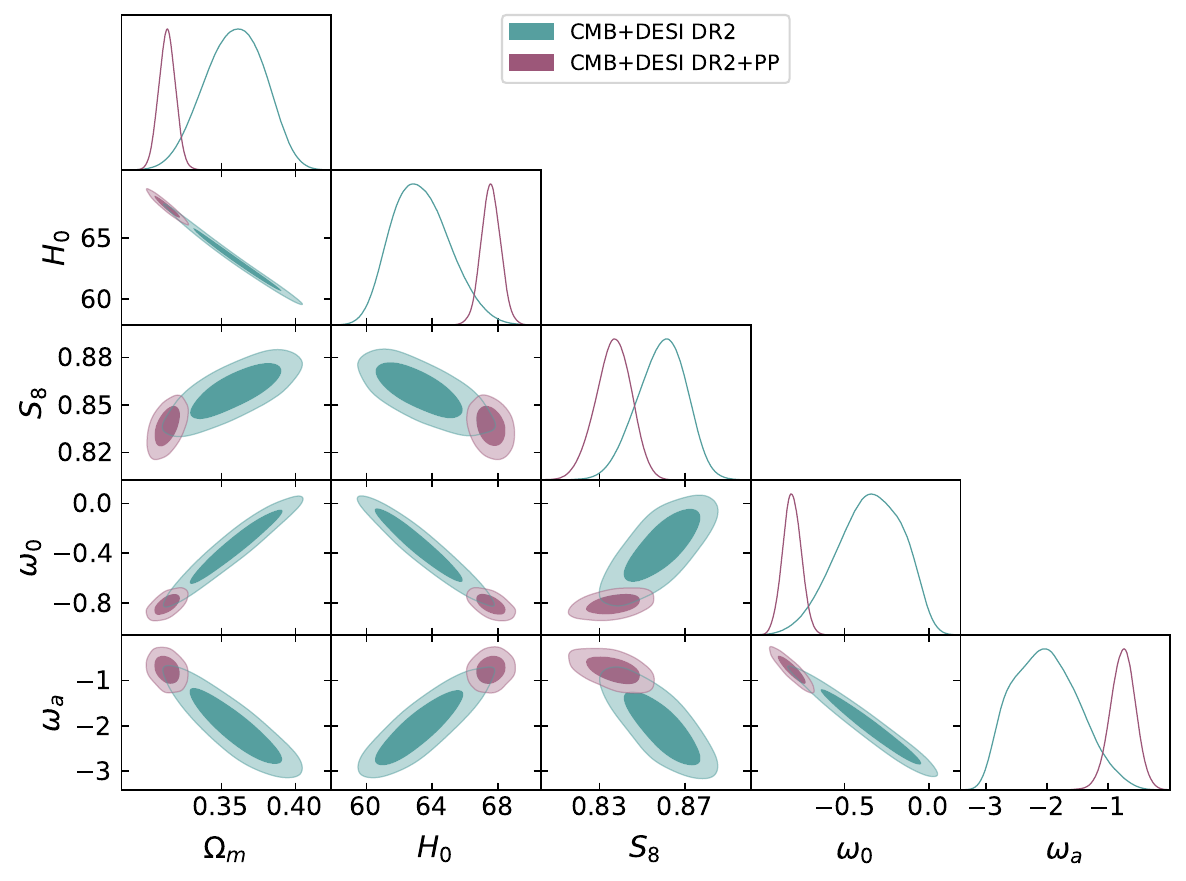}
    \label{fig:cpl_triangle}
}

\caption{Triangular plots showing the marginalized posterior distributions and parameter degeneracies for (a) the QDEE model and (b) the CPL parametrization. Contours correspond to the CMB+DESI DR2 (teal) and CMB+DESI DR2+PP (purple) datasets.}
\label{fig:model_triangles}
\end{figure*}

A key distinction emerges in the inferred values of the Hubble constant $H_0$. Under CMB-only constraints, QDEE allows substantially larger values ($H_0 = 71.80^{+3.2}_{-1.9}~\mathrm{km/s/Mpc}$) than $\Lambda\mathrm{CDM}$ ($H_0=67.08\pm0.40~\mathrm{km/s/Mpc}$), partially alleviating the Hubble tension without invoking early-Universe modifications. This shift is driven by the negative $\Omega_1$ contribution and small positive $\Omega_2$, which collectively modify the late-time expansion history. A negative correlation between $H_0$ and $\Omega_1$ parameter is clearly visible in Fig.~\ref{fig:taylor_triangle}, consistent with the role of $\Omega_1$ in modifying the late-time expansion rate. Fig. \ref{fig:taylor_triangle}, also shows that the inclusion of the PP dataset (purple contours) visibly shrinks the posteriors, particularly for $H_0$ and $\Omega_{\rm m}$, reinforcing the constraining power of low-redshift distance measurements. Thus, once DESI DR2 and PP datasets are included, the preferred QDEE value gradually shifts downward towards $68.72\pm0.51~\mathrm{km/s/Mpc}$, indicating that low-redshift geometric probes strongly constrain deviations from $\Lambda\mathrm{CDM}$ and pull the model closer to standard expansion. Nevertheless, QDEE consistently maintains slightly higher $H_0$ values than $\Lambda\mathrm{CDM}$, suggesting modest tension relief while preserving observational consistency.

On the other hand, Fig.~\ref{fig:cpl_triangle} reveals that the CPL parametrization shows considerably broader and more elongated contours, especially for $H_0$, $w_0$, and $w_a$, indicative of stronger parameter degeneracies and poor constraining power from CMB data alone. As a result, the model exhibits significantly less stable behavior than both QDEE and $\Lambda\mathrm{CDM}$. It is worth noting that the broad degeneracies exhibited by the CPL parametrization, particularly under CMB-only constraints, are a well-known feature of EoS expansions with weak late-time geometric information. In this sense, the larger parameter volume and extended degeneracy directions allowed within CPL naturally produce less stable cosmological constraints. Our present comparison indicates the relative phenomenological stability of density-based versus EoS parametrizations under the adopted observational combinations, rather than establishing a fundamental inconsistency of the CPL parametrization itself.

For CMB-only data, CPL allows extremely large values of the Hubble constant ($H_0 \sim 92.01~\mathrm{km/s/Mpc}$), as shown in Table~\ref{constraints}, with correspondingly large uncertainties, reflecting the well-known inability of this framework to tightly constrain late-time expansion using early Universe information alone. Moreover, QDEE produces a significantly more moderate and observationally consistent shift along the same degeneracy direction. For the CMB+DESI DR2 combination, QDEE favors $H_0 \simeq 70.07 \pm 0.73~\mathrm{km/s/Mpc}$, corresponding to a $2.3\sigma$ tension with SH0ES together with $\Omega_{\rm m} \simeq 0.2909 \pm 0.0058$. This value partially alleviates the Hubble tension while maintaining a physically plausible matter density. Standard $\Lambda\mathrm{CDM}$ yields an intermediate result, with $H_0 \simeq 68.10 \pm 0.27~\mathrm{km/s/Mpc}$, corresponding to a tension of approximately $4.6\sigma$ with SH0ES.

\begin{figure}[!hbt]
    \includegraphics[width=\columnwidth]{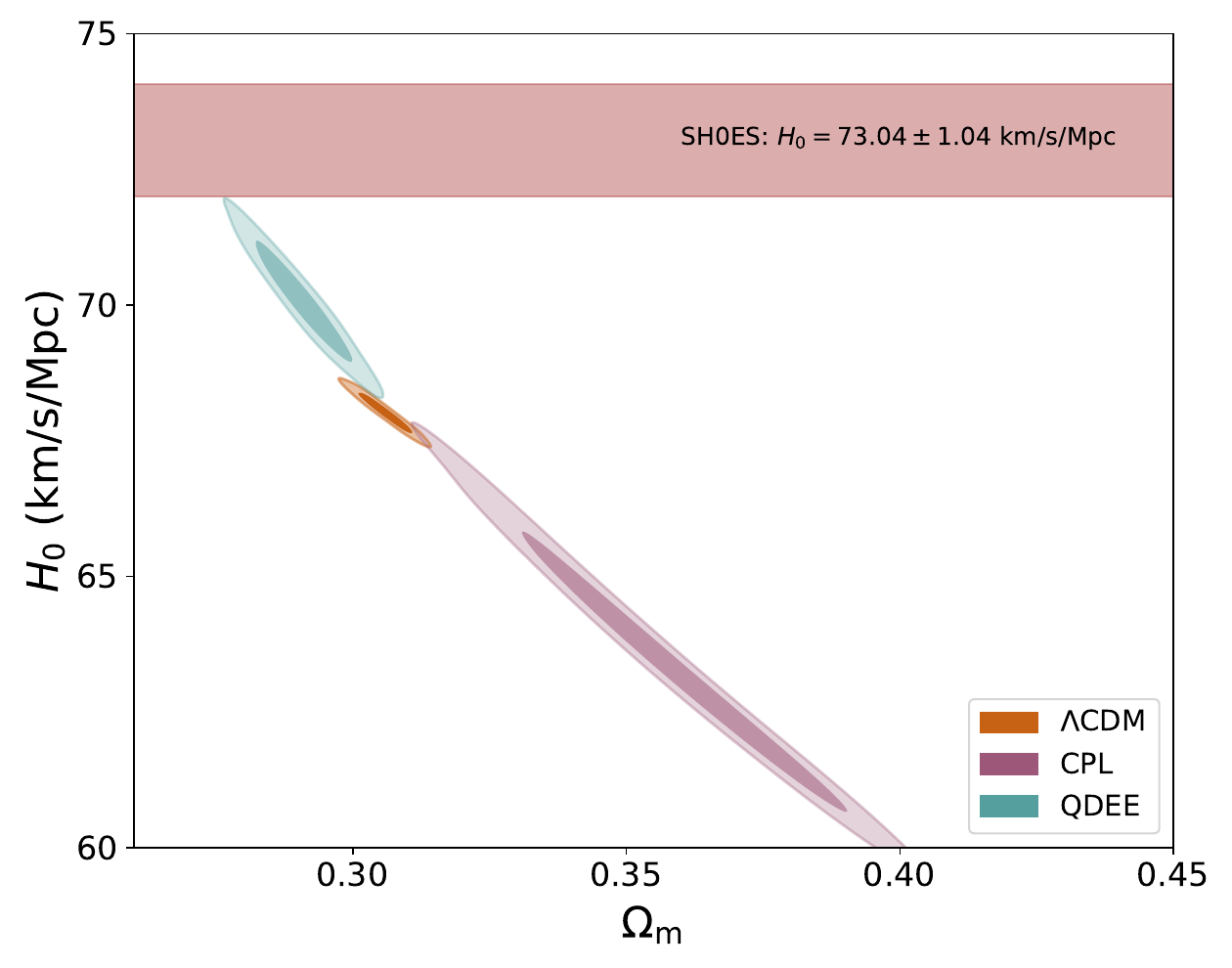}
    \caption{Joint constraints in the $H_0$–$\Omega_{\rm m}$ plane for the $\Lambda\mathrm{CDM}$ (orange), CPL (purple), and QDEE (teal) models. The brown horizontal band corresponds to the SH0ES measurement of $H_0$, respectively.}
    \label{H0plot}
\end{figure}

Fig. \ref{H0plot} shows that all the three cosmological models exhibit the expected negative correlation between $H_0$ and $\Omega_{\rm m}$, reflecting the well-known geometric degeneracy present in the combined CMB+DESI DR2 constraints. However, each DE parametrization shifts the posterior along this degeneracy direction in physically distinct ways. The CPL framework favors a substantially displaced region of parameter space, yielding a notably low Hubble constant of $H_0 \simeq 63.3^{+1.5}_{-2.0}~\mathrm{km/s/Mpc}$ and a correspondingly high matter density of $\Omega_{\rm m} \simeq 0.359^{+0.022}_{-0.019}$, corresponding to a $\sim 5.3\sigma$ tension with the SH0ES measurement $H_0 = 73.04 \pm 1.04~\mathrm{km/s/Mpc}$ \cite{Riess_2022}. Such behavior indicates that, when constrained by DESI DR2, the CPL parametrization effectively drives the solution toward an observationally disfavored region of parameter space. In contrast, the QDEE model shifts the posterior towards moderately higher values of $H_0$, providing a more balanced modification of the late-time expansion history while maintaining physically reasonable matter densities. Standard $\Lambda\mathrm{CDM}$ occupies an intermediate position, with $H_0 \simeq 68.10 \pm 0.27~\mathrm{km/s/Mpc}$, corresponding to $\sim 3.2\sigma$ tension with SH0ES. These results highlight that, although all models are subject to the same underlying degeneracy, physically motivated density-based parametrizations such as QDEE yield more stable and observationally consistent cosmological solutions compared to purely phenomenological equation-of-state extensions.

The inclusion of PP supernova data does not substantially change the overall behavior of the models, but it further tightens late-time geometric constraints. For both $\Lambda\mathrm{CDM}$ and QDEE, the inferred values of $H_0$ shift only mildly downward, indicating that these models remain relatively stable once CMB and DESI DR2 constraints are already included. In contrast, CPL shows a more pronounced response: its preferred value increases significantly from $H_0 \simeq 63.30^{+1.5}_{-2.0}~\mathrm{km/s/Mpc}$ to $67.58 \pm 0.59~\mathrm{km/s/Mpc}$, reflecting the stronger role of supernova data in reducing CPL's broader parameter degeneracies. Despite these adjustments, all models remain in tension with the SH0ES local measurement, implying that late-time DE modifications alone are insufficient to fully resolve the current $H_0$ discrepancy. We also point out that although the QDEE framework systematically shifts the inferred value of the Hubble constant toward higher values relative to $\Lambda\mathrm{CDM}$, the resulting alleviation of the Hubble tension remains moderate once the full dataset combination is considered. In particular, the inclusion of DESI DR2 and PP data significantly constrains the allowed late-time deviations from a pure cosmological constant, pulling the expansion history closer to the standard concordance scenario. As a result, while QDEE provides a mild reduction of the $H_0$ discrepancy without modifying early-Universe physics, the present results do not indicate a complete resolution of the tension.

\begin{figure*}[t]
\centering

\subfigure[]{
    \includegraphics[width=0.48\textwidth]{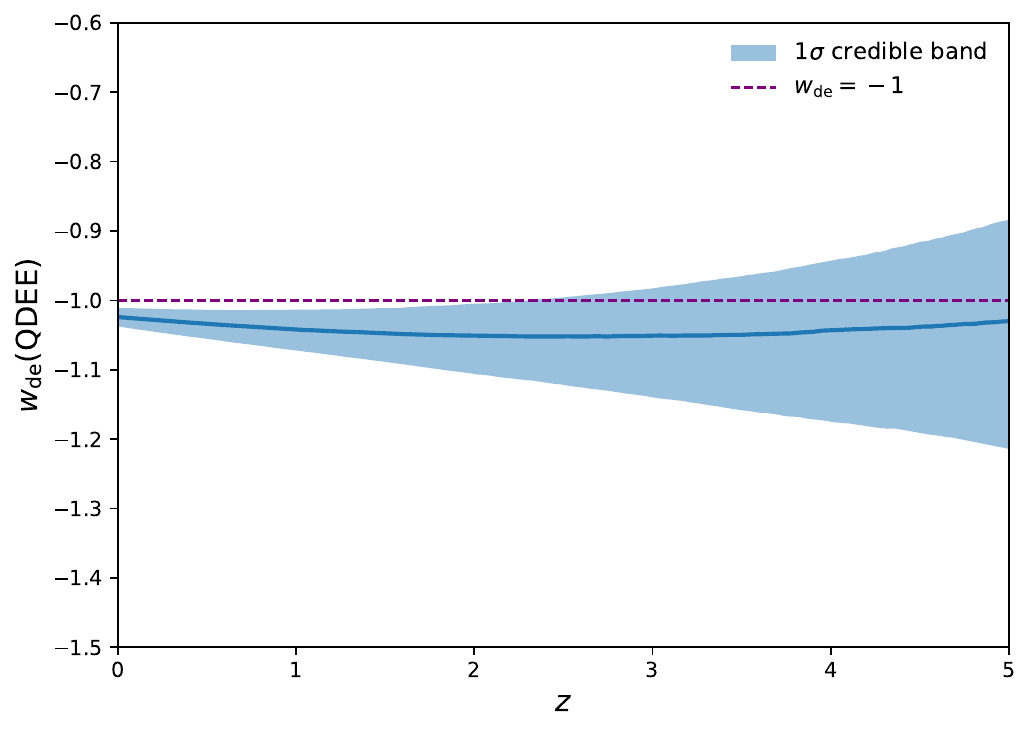}
    \label{QDEE}
}
\hfill
\subfigure[]{
    \includegraphics[width=0.48\textwidth]{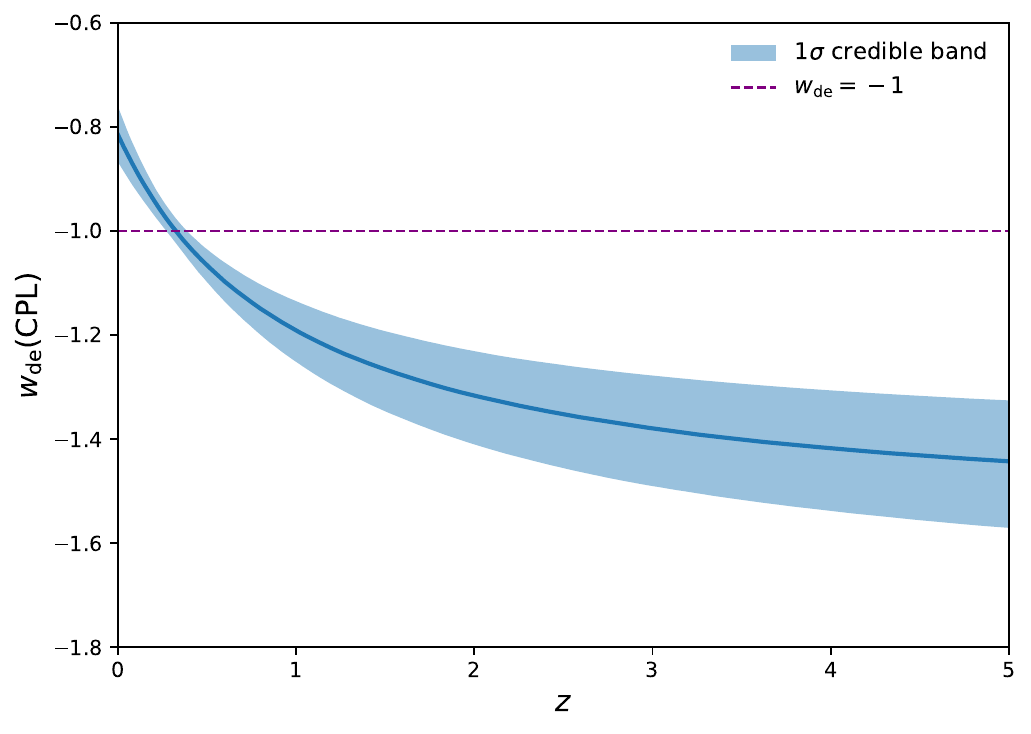}
    \label{CPL}
}
\caption{Reconstructed equation of state $w_{\rm de}(z)$ for the QDEE and  CPL models from CMB+DESI DR2+PP data. The shaded region represents the $1\sigma$ confidence interval. The dashed purple line indicates the phantom divide at $w_{\rm de} = -1$.}
\label{fig:reconstruction_plots}
\end{figure*}

In Fig.~\ref{QDEE}, we have that the QDEE EoS reconstruction remains close to the cosmological constant value $w_{\rm de} = -1$ over the redshift range $0 \leq z \leq 5$. The posterior mean lies mildly in the phantom regime, while the $1\sigma$ credible band broadens toward higher redshifts, indicating that deviations from $w_{\rm de} = -1$ are only weakly constrained at earlier epochs. Thus, the QDEE model exhibits $\Lambda$CDM-like behaviour with a mild preference for phantom-like evolution. In contrast, Fig.~\ref{CPL} shows a more pronounced redshift evolution. The reconstructed EoS begins in the phantom regime ($w_{\rm de} < -1$) at high redshift, crosses the phantom divide, and evolves toward significantly less negative values ($w_{\rm de} > -1$) at lower redshifts. This stronger evolution reflects the greater dynamical flexibility of the CPL parametrization compared with the QDEE model.

\subsection{Bayesian Evidence and Model Comparison}
To evaluate the relative goodness of fit of the cosmological models considered in this work, we examine the relative minimum chi-square statistic, defined as $\Delta\chi^2_{\rm min} = \chi^2_{\rm min, model} - \chi^2_{\rm min,\Lambda\rm CDM}$. For a Gaussian likelihood, the minimum chi-square is related to the likelihood through $\chi^2_{\rm min} = -2~ \rm ln\mathcal{L}$, where $\mathcal{L}$ denotes the likelihood function evaluated at the best-fit cosmological parameters \cite{Cowan:1998ji, Trotta:2008qt}. The values reported in Table~\ref{constraints} show that among the extended models, CPL generally yields the largest improvement in fit, reflected in the most negative $\Delta \chi^2_{\rm min}$, followed by QDEE, while standard $\Lambda\mathrm {CDM}$ corresponds to the reference case with $\Delta \chi^2_{\rm min}=0$.

\begin{figure}[hbt]
    \centering
    \includegraphics[width=\columnwidth]{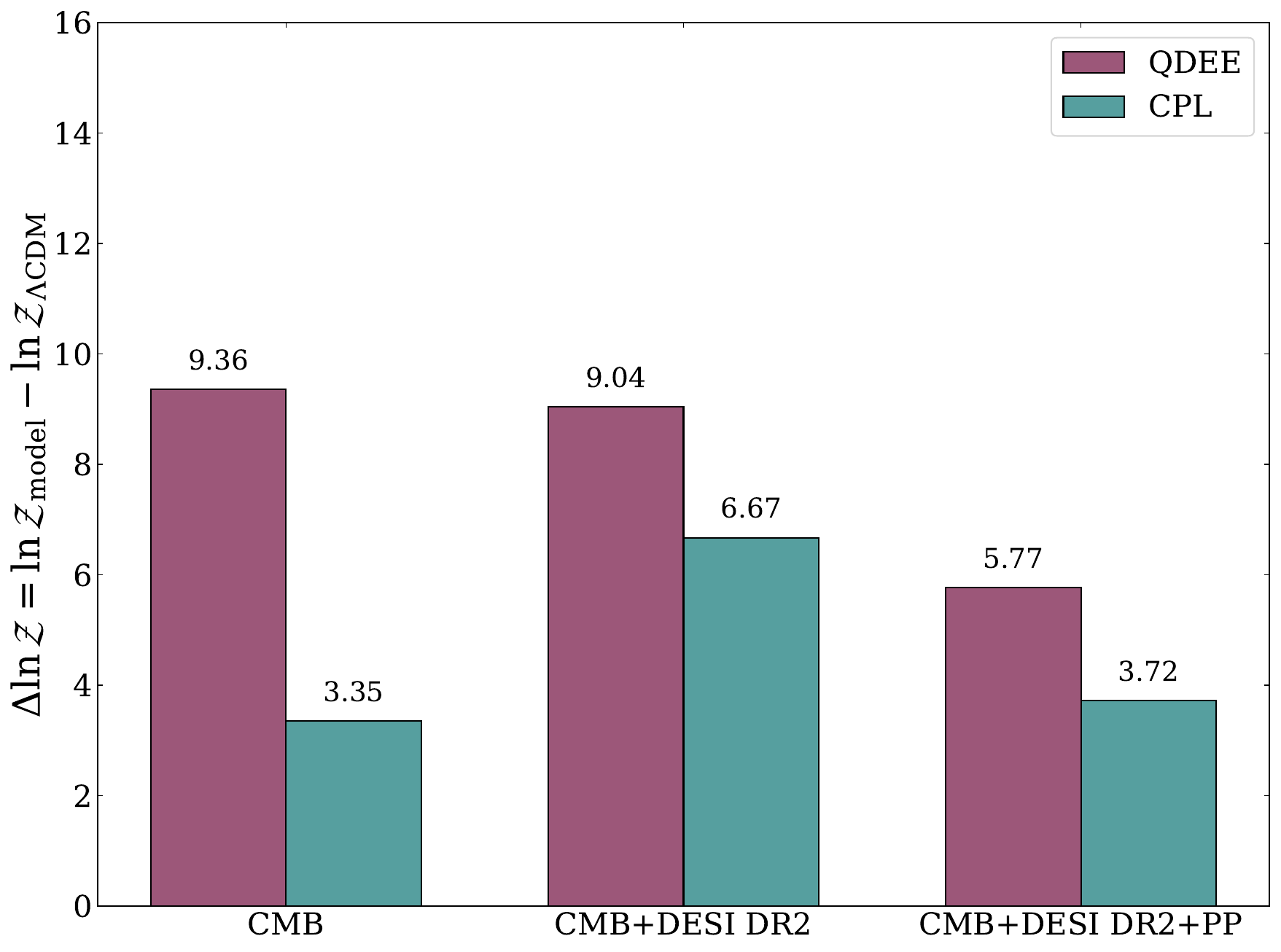}
    \caption{Log-Bayesian evidence of QDEE and CPL models relative to $\Lambda$CDM for different datasets.}
    \label{BE}
\end{figure}

However, the minimum chi-square alone assesses model performance only at a single point in parameter space and does not account for model complexity or the volume of viable parameter space. Consequently, it is insufficient as a standalone criterion for robust model comparison. A more comprehensive statistical measure is provided by the Bayesian evidence, which naturally balances goodness of fit against parameter-space volume and thus penalizes unnecessarily flexible models.

In this work, the Bayesian evidence is computed using the \texttt{MCEvidence} package \cite{Heavens:2017afc}, which estimates log-Bayesian evidence $\ln\mathcal{Z}$ directly from MCMC samples through a $k^{\rm th}$-nearest-neighbor algorithm. An initial burn-in fraction of 30\% of each chain is removed to ensure stable and unbiased evidence estimates. Model comparison is then performed using Bayes factor $\ln\mathcal{B}_{\rm model,\Lambda \mathrm{CDM}}\equiv\Delta\ln \mathcal{Z}$, defined as the difference in log-Bayesian evidence relative to the reference $\Lambda\mathrm{CDM}$ model, $\Delta\ln \mathcal{Z} = \ln\mathcal{Z}_{\rm model} - \ln \mathcal{Z}_{\Lambda{\mathrm{CDM}}}$. The statistical significance of these differences is interpreted according to Jeffreys' scale \cite{Kass:1995loi, Trotta:2008qt, Gordon:2007xm}. Specifically, values in the range $0\leq |\Delta\ln\mathcal{Z}| < 1$ are regarded as inconclusive, indicating no statistically meaningful preference between models. A weak level of evidence corresponds to $1 \leq |\Delta \ln\mathcal{Z}| < 2.5$, while moderate evidence is characterized by $2.5 \leq |\Delta\ln \mathcal{Z}| < 5$. Strong evidence is associated with $5 \leq |\Delta\ln\mathcal{Z}| < 10$, and values satisfying $|\Delta\ln\mathcal{Z}| \geq 10$ are interpreted as very strong evidence in favor of one model over another.

The Bayesian evidence results shown in Fig.~\ref{BE} and Table~\ref{constraints} reveal that the QDEE model is consistently favored over standard $\Lambda\mathrm{CDM}$ across multiple dataset combinations. In particular, we find $\Delta\ln\mathcal{Z} \simeq 9$ approximately for both the CMB-only and CMB+DESI DR2 analyses, and $\Delta\ln\mathcal{Z} \simeq 5.77$ for the full CMB+DESI DR2+PP dataset. According to Jeffreys' scale, these values correspond to strong evidence in favor of the QDEE framework. Typically, while both dynamical DE extensions improve the raw fit to current cosmological observations relative to $\Lambda\mathrm{CDM}$, the Bayesian analysis demonstrates that the QDEE model provides the most statistically efficient improvement. These findings suggest that physically constrained late-time DE modifications offer a more compelling description of current cosmological data than either the standard cosmological constant or less restricted phenomenological parametrizations such as CPL.

It is worth noting that the reported values of $\Delta\ln\mathcal{Z}$ may exhibit some sensitivity to the adopted prior volume and sampling strategy. In particular, phenomenological late-time dark-energy extensions can present nontrivial parameter degeneracies and mildly non-Gaussian posterior structures, which may affect evidence reconstruction from MCMC chains. Although our analysis employs physically motivated priors to ensure a stable cosmological evolution, a more exhaustive investigation of prior dependence, including broader prior intervals and independent nested-sampling analyses, would be valuable for establishing the quantitative stability of the inferred evidence preference over $\Lambda\mathrm{CDM}$.

\subsection{Posterior Predictive Checks}
\begin{figure*}[t]
\centering

\subfigure[~\texorpdfstring{$\Lambda$CDM}{LCDM}]{
    \includegraphics[width=5.65cm]{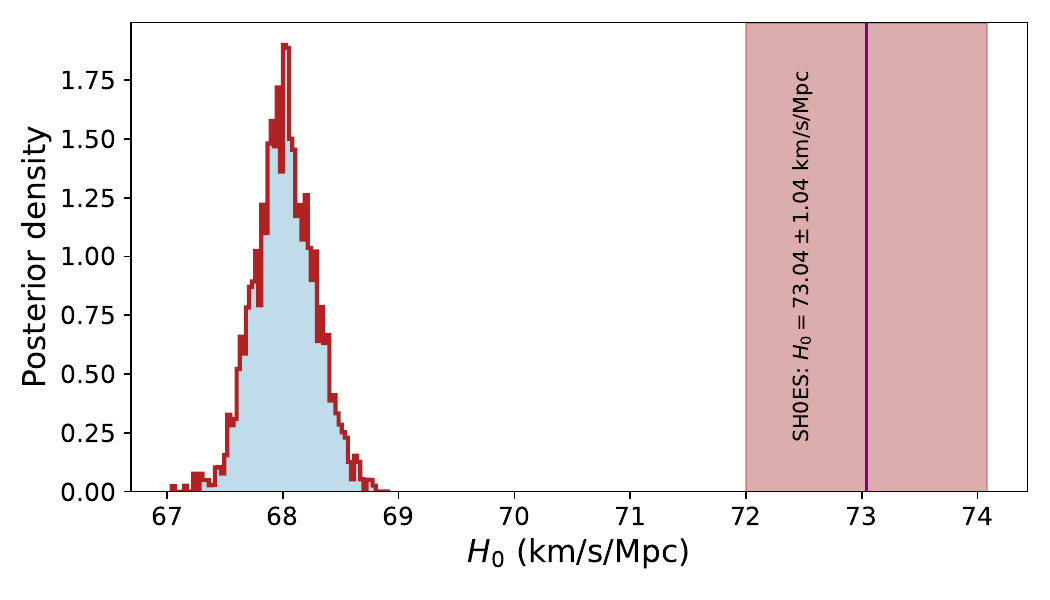}
    \label{lcdm_h0}
}
\hfill
\subfigure[~CPL]{
    \includegraphics[width=5.65cm]{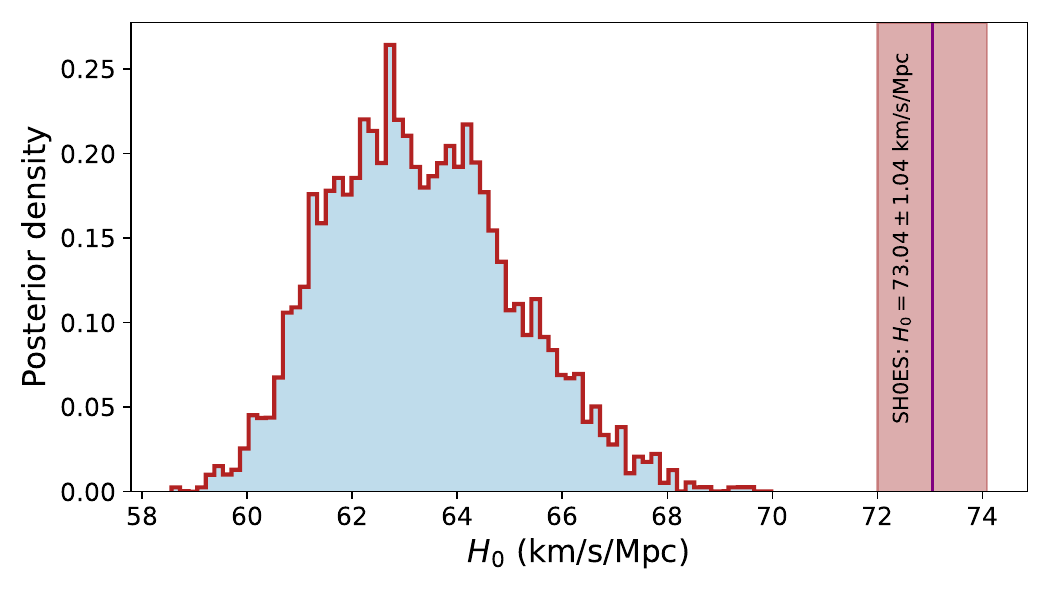}
    \label{cpl_h0}
}
\hfill
\subfigure[~QDEE]{
    \includegraphics[width=5.65cm]{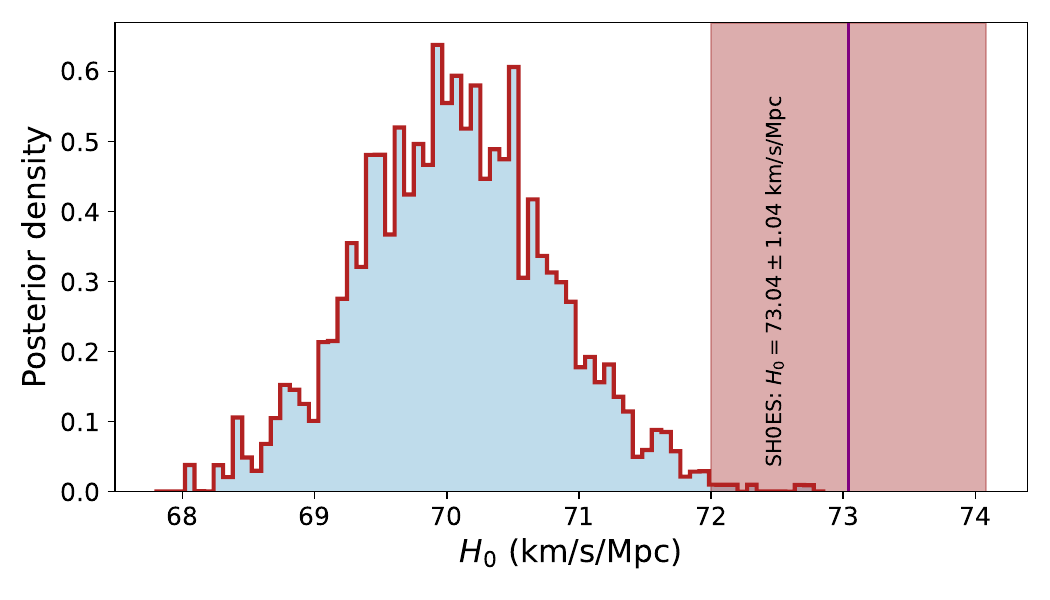} 
    \label{qdee_h0}
}
\caption{Marginalized posterior distribution of $H_0$ for the $\Lambda\mathrm{CDM}$, CPL and QDEE models obtained from the CMB+DESI dataset. The shaded vertical bands represent the SH0ES measurements.}
\label{fig:cpl_h0}
\end{figure*}

In addition to the parameter constraints previously discussed, we examine the predictive consistency of the three cosmological models using PPC \cite{GelmanRubin1992, BSLYNCH2005135, Meng1994PPP, 2024arXiv241215809S, rubin2014posterior}. Unlike traditional goodness-of-fit statistics, PPC tests whether a model that is calibrated using a given dataset can successfully predict independent observables. The posterior predictive distribution (PPD) provides a way to assess how close the replicated data drawn from the PPD resemble the observed data. If the data follows the pattern closely, it indicates we have chosen our model well. For a full Bayesian framework $p(d, \theta)$, the model posterior for observed data $d_{\rm obs}$ is $P(\theta\mid d_{\rm obs})$. The posterior predictive density for a replicated/new data $d_{\rm rep}$ is then\footnote{See~\cite{Gelman2003, DES:2020lei}; the prior operation $I$ is dropped for conciseness.}
\begin{equation} 
P(d_{\rm rep}\mid d_{\rm obs}) = \int P(d_{\rm rep}\mid \theta)P(\theta\mid d_{\rm obs})\mathrm{d}\theta,
\end{equation}
where the conditional independence of $d_{\rm rep}$ from $d_{\rm obs}$ given $\theta$ has been used to simplify the integrand. The integral above marginalizes over the parameters $\theta$, leaving $p(d_{\rm rep}\mid d_{\rm obs})$ of future observations given past observations. This distribution encodes all the observed data and tells us about the future replicated data, making it a self-consistent tool for PPC. 

To quantify the comparison between the observed and replicated datasets, test statistics $T(d_{\rm obs})$ and $T(d_{\rm rep})$ have been defined as a function of $\theta$, which capture relevant features of the observed dataset and evaluate the same statistic $T(d_{\rm rep})$ on the replicated data. In particular, for Gaussian-distributed data, a common choice is $\chi^2$, i.e.,

\begin{equation}
    T(d_{i},\theta) = (d_{i}-\mu(\theta))^{T}C^{-1}(d_{i}-\mu(\theta)),
\end{equation}
where $i \in \{\text{obs}, \text{rep}\}$, $\mu(\theta)$ is the model evaluated at parameter values $\theta$, and $C$ is the covariance matrix. Using $\chi^2$ as a test statistic can bias the p-value if the two experiments considered are disjoint subsets of a data vector, which constrain very different volumes of the parameter volume \cite{DES:2020lei}. This statistic should therefore be interpreted with caution in such cases. Another test statistic used in this work is the max discrepancy statistic \cite{Gelman:1996ppc, Gelman2003} defined as:
\begin{equation}
T(d_i, \theta)=\max_n |y_{n,i}|,
\end{equation}
where $y_{n,i}=\left[L^{-1}(d_{i}-\mu(\theta))\right]_n$ are the whitened residuals. In this equation, $i$ and $\mu(\theta)$ are the same as above, $n\in\{1, \ldots, N\}$ is the index running over the $N$ data points in dataset $i$, and $L$ is the Cholesky factor of the covariance matrix $C$. The test statistic for $\chi^2$ is represented as $T^{\chi^2}_{\rm obs}$ and $T^{\chi^2}_{\rm rep}$ whereas the test statistic for max discrepancy is represented as $T^{\rm max}_{\rm obs}$ and $T^{\rm max}_{\rm rep}$ throughout the text. The statistic quantifies whether the dataset contains an observation that is too extreme to plausibly come from the model.

The classical posterior predictive p-value is then given by 
\begin{equation}
      p = P\!\left[T(d_{\rm rep},\theta) \geq T(d_{\rm obs},\theta)\mid d_{\rm obs}\right].
\end{equation}
A $p$-value approaching 0 or 1 may signal tension between the model predictions and the observed data, whereas values near the centre of the unit interval indicate satisfactory consistency \cite{Gelman:1996ppc}. 

\begin{figure*}[t]
\centering

\subfigure[~\texorpdfstring{$\Lambda$CDM}{LCDM}]{
    \includegraphics[width=5.65cm]{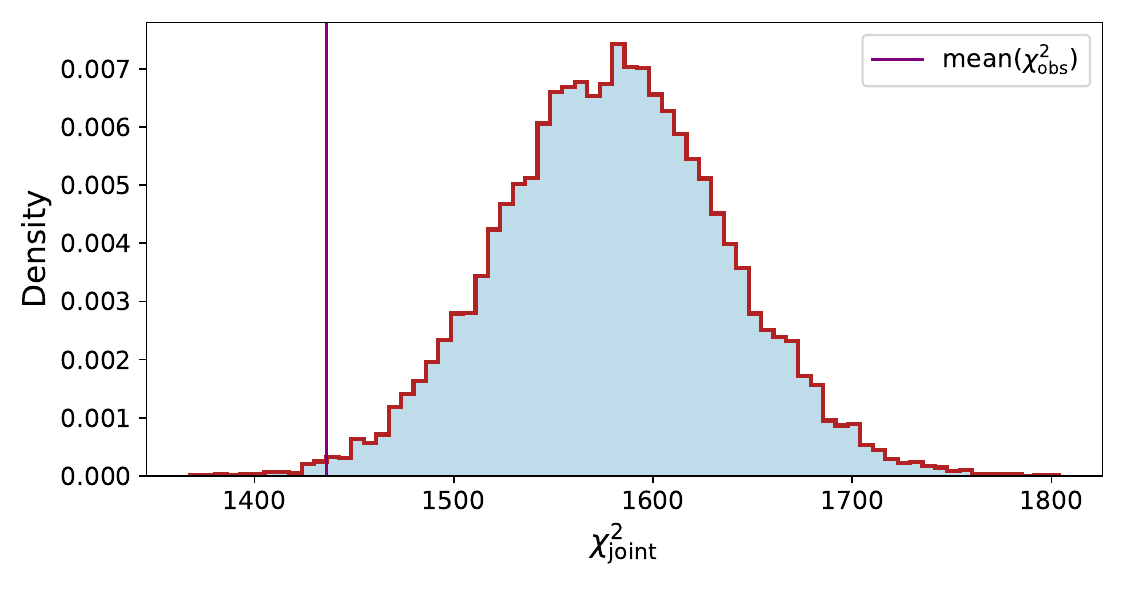}
    \label{lcdm_chi2}
}
\hfill
\subfigure[~CPL]{
    \includegraphics[width=5.65cm]{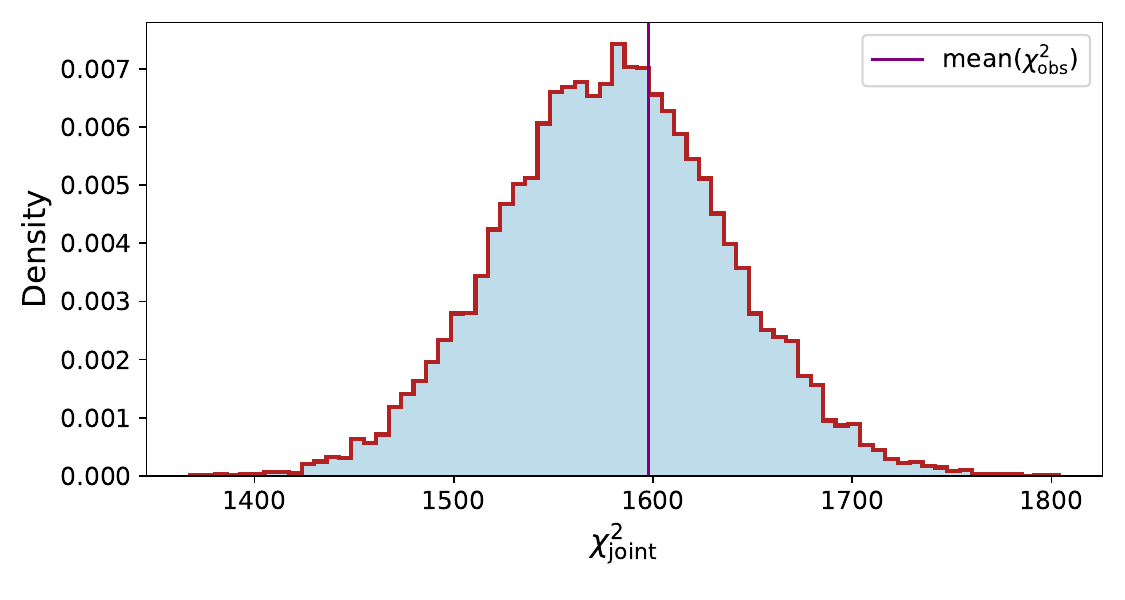}
    \label{cpl_chi2}
}
\hfill
\subfigure[~QDEE]{
    \includegraphics[width=5.65cm]{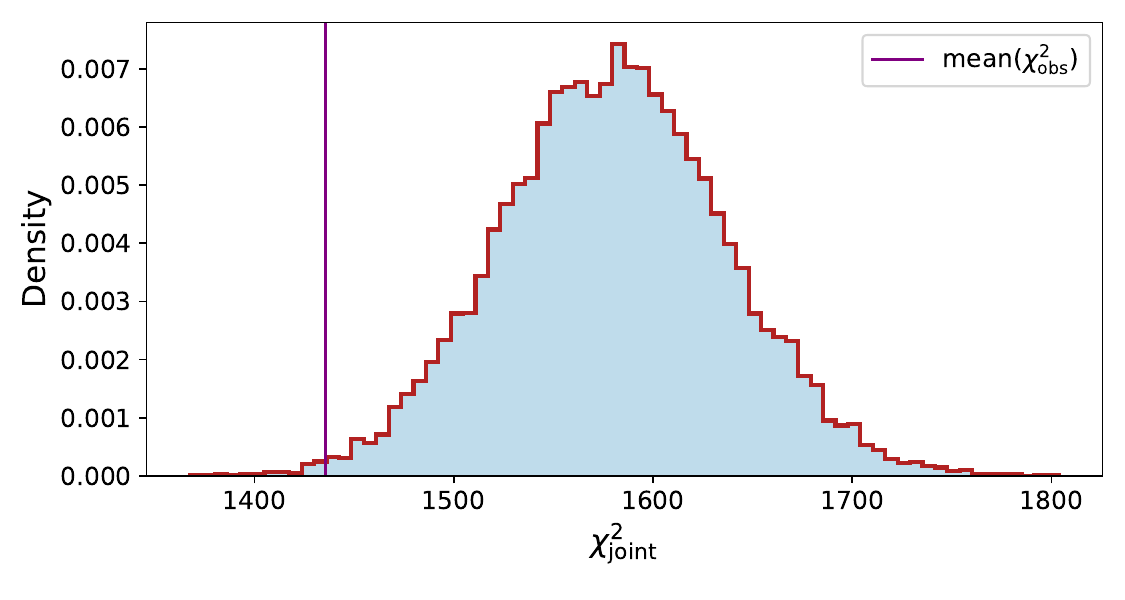} 
    \label{qdee_chi2}
}
\caption{Posterior predictive distribution of the global $\chi^2_{\rm rep}$ statistic for the $\Lambda\mathrm{CDM}$, CPL and QDEE models using the CMB+DESI dataset. The vertical line indicates the observed value, while the distribution represents replicated realizations drawn from the posterior.}
\label{fig:cpl_chi2}
\end{figure*}

\begin{table}[hbp]
\centering
\caption{PPC diagnostics for the three cosmological models using the CMB+DESI DR2 dataset.}
\begin{tabular}{lccc}
\hline\hline
Statistic & QDEE & CPL & $\Lambda\mathrm{CDM}$ \\
\hline
\vspace{0.1 cm}
$T^{\chi^2}_{\rm obs}$ 
& 1435.67 
& 1597.56 
& 1436.17 \\
\hline
\vspace{0.1 cm}
$ T^{\chi^2}_{\rm rep}$
& 1580.25 
& 1580.25  
& 1580.25   \\ 
\hline
\vspace{0.1 cm}
$p_{\rm PPC}^{\chi^2}$
& 0.9953 
& 0.458 
& 0.99545  \\
\hline
\vspace{0.1 cm}
$T^{\rm max}_{\rm obs}$ 
& 4.15 
& 9.35
& 4.83 \\
\hline
\vspace{0.1 cm}
$T^{\rm max}_{\rm rep}$ 
& 3.56 
& 3.56 
& 3.56 \\
\hline
\vspace{0.1 cm}
$p_{\rm PPC}^{\max}$
& 0.054
& $2\times10^{-4}$ 
& $5\times10^{-3}$  \\ 
\hline
\vspace{0.1 cm}
$\langle H_0^{\rm rep}\rangle$
& 70.07 
& 63.31  
& 68.01  \\ 
\hline\hline
\end{tabular}
\label{tab:ppc_results}
\end{table}

The primary PPC analysis is performed using the CMB+DESI DR2 chains in order to test the predictive consistency of late-time distance observables, in particular the supernova distance modulus ($\mu$) and the Hubble constant ($H_0$), without conditioning directly on the supernova likelihood. Table \ref{tab:ppc_results} reports: the observed chi-square statistic $T^{\chi^2}_{\rm obs}$; the mean replicated chi-square $\langle T^{\chi^2}_{\rm rep} \rangle$ obtained from posterior predictive realizations; the posterior predictive $p$-value ${p}_{\rm PPC}$, defined as the fraction of replicated statistics exceeding the observed value; the maximum observed discrepancy statistic $T^{\rm max}_{\rm obs}$; the maximum replicated discrepancy statistic $T^{\rm max}_{\rm rep}$; the posterior predictive probability of maximum discrepancy statistic $p_{\rm PPC}^{\max}$; and the mean predicted Hubble parameter $\langle H_0^{\rm rep} \rangle$ from the replicated samples\footnote{Note that $T^{\chi^2}_{\rm rep}$ and $T^{\rm max}_{\rm rep}$ are identical across models since the replicated data are drawn from the same PP covariance structure.}.

\begin{figure*}[t]
\centering

\subfigure[~\texorpdfstring{$\Lambda$CDM}{LCDM}]{
    \includegraphics[width=5.65cm]{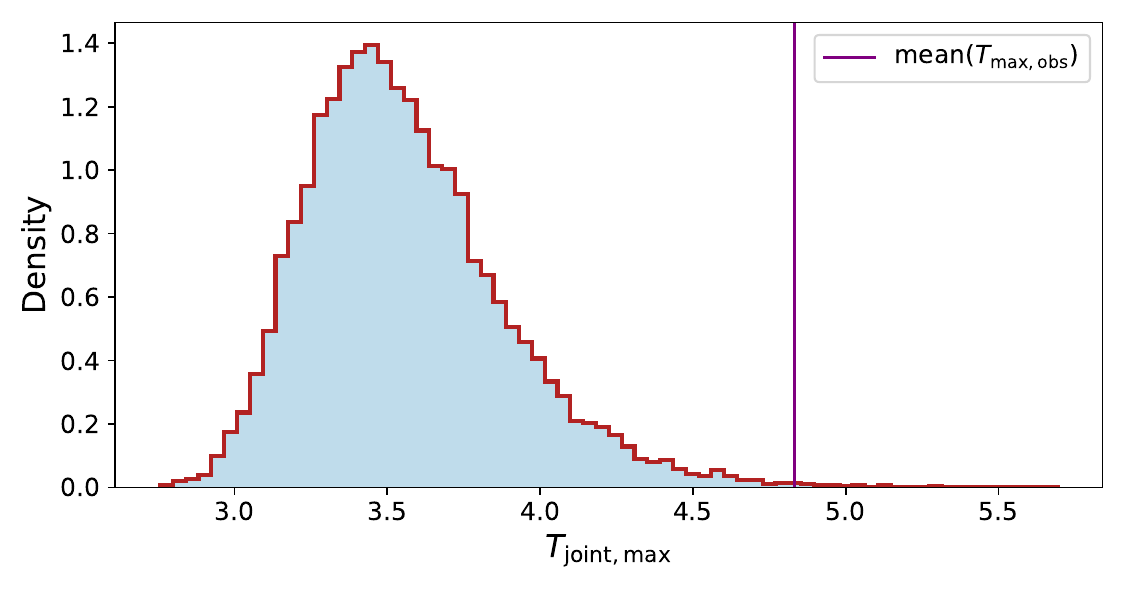}
    \label{lcdm_max}
}
\hfill
\subfigure[~CPL]{
    \includegraphics[width=5.65cm]{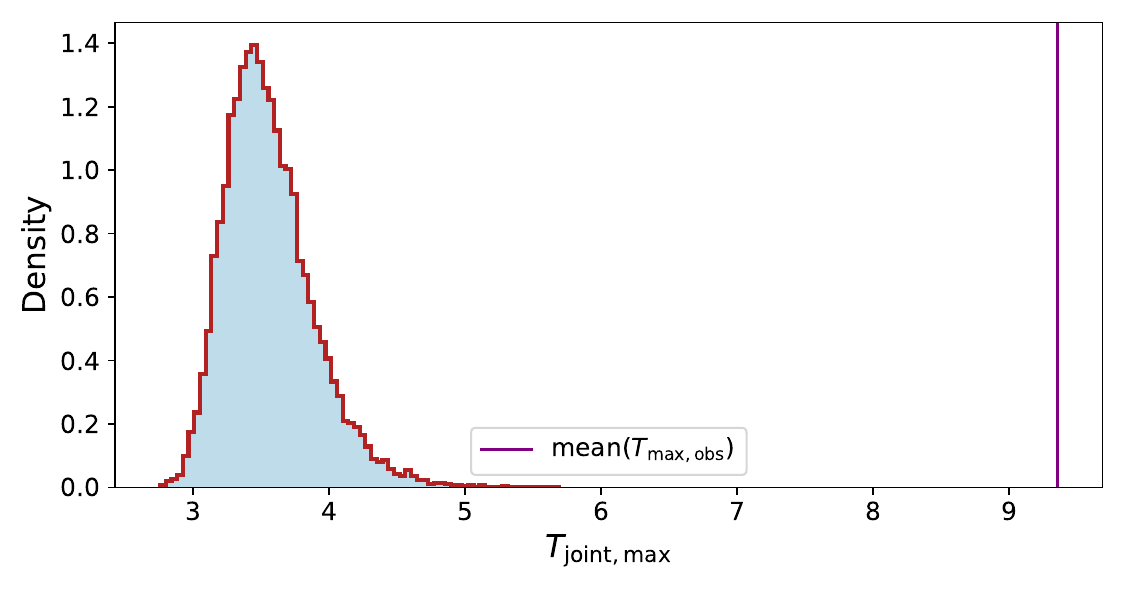}
    \label{cpl_max}
}
\hfill
\subfigure[~QDEE]{
    \includegraphics[width=5.65cm]{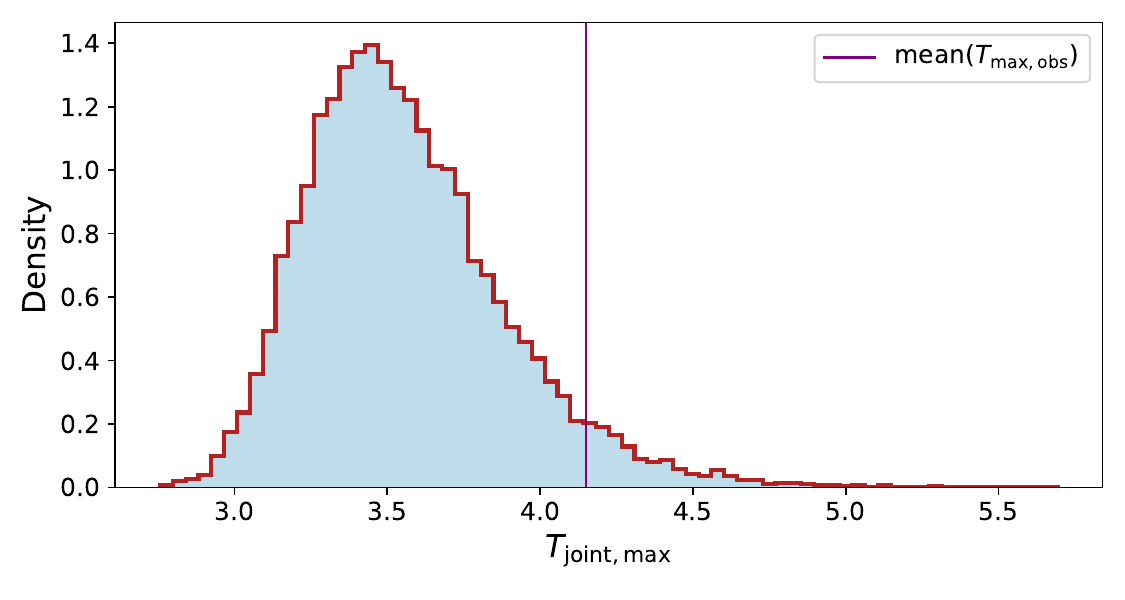}
    \label{qdee_max}
}
\caption{Posterior predictive distribution of the maximum discrepancy statistic $T_{\max}$ for the $\Lambda\mathrm{CDM}$ model, CPL parametrization case and the QDEE model using the CMB+DESI dataset. The vertical line indicates the observed value, while the histogram shows replicated realizations drawn from the posterior.}
\label{fig:cpl_max}
\end{figure*}

For $\Lambda\mathrm{CDM}$, we obtain $H_0 = 68.01~\mathrm{km/s/Mpc}$, consistent with the conventional cosmological expectation, although still noticeably lower than the local SH0ES determination as shown in Fig.~\ref{lcdm_h0} However, the CPL parametrization yields a substantially lower value of $H_0 = 63.30~\mathrm{km/s/Mpc}$, placing it in even stronger tension with local distance-ladder measurements and significantly farther from the SH0ES constraint, as shown in Fig.~\ref{cpl_h0}. The QDEE model yields $H_0 = 70.07~\mathrm{km/s/Mpc}$, representing a moderate upward shift relative to $\Lambda\mathrm{CDM}$ and moving somewhat closer to the SH0ES preferred region, as illustrated in Fig.~\ref{qdee_h0}. Thus, QDEE provides partial alleviation of the Hubble tension without requiring the extreme parameter shifts or DE behavior often associated with more flexible phenomenological parametrizations such as CPL.

From Fig.~\ref{lcdm_chi2} and Fig.~\ref{qdee_chi2}, we observe that both the $\Lambda\mathrm{CDM}$ and QDEE models display similar behavior. As also shown in Table~\ref{tab:ppc_results}, both models yield $p_{\rm PPC} \sim 0.995$, indicating that the 
observed $\chi^2$ is unusually small compared with the replicated distribution. However, in Fig.~\ref{cpl_chi2}, the observed $\chi^2$ value lies near the center of the posterior predictive distribution, indicating an overall acceptable goodness of fit of the CPL model. 

A more informative diagnostic is provided by the maximum discrepancy statistic as a complementary tool for the posterior predictive test. For the CPL model, the observed value of the maximum discrepancy statistic, $T_{\rm obs}^{\rm{max}} \sim 9.35$ substantially exceeds the typical replicated value, $T_{\rm rep}^{\rm max} \simeq 3.56$, resulting in $p_{\rm PPC}^{\rm max} = 2\times10^{-4}$, indicating that at least one subset of the data is poorly described. 

On the other hand, both $\Lambda\mathrm{CDM}$ and QDEE yield substantially more acceptable values of $T_{\rm obs}^{\rm max}$ with respect to $T^{\rm max}_{\rm rep}$ as shown in Table~\ref{tab:ppc_results}. The same behaviour can be seen in Figs.~\ref{lcdm_max},~\ref{cpl_max} and \ref{qdee_max}. While these values indicate mild residual tension for both models, they remain within an acceptable statistical range and suggest that QDEE provides a comparatively coherent description across datasets. Similarly, we have the same situation for the $\Lambda\mathrm{CDM}$ performance that reinforces its internal consistency despite its known tensions in global parameter inference.

We stress that the PPC analysis indicates that the inferred consistency of the cosmological models depends sensitively on the diagnostic employed. The global $\chi^2$-based PPC yields values close to unity for both $\Lambda\mathrm{CDM}$ and the QDEE model, reflecting that the observed residuals are smaller than typical replicated realizations. While such high $p_{\rm PPC}$ values do not signal a poor fit, they also suggest that the global $\chi^2$ statistic may not be sufficiently discriminating in this context, as it averages residuals over a large number of data points. Conversely, the CPL parametrization produces a value $p_{\rm PPC} \simeq 0.46$, which is closer to the ideal expectation, indicating an adequate description of the overall variance of the data. Thus, our results highlight that a good global fit does not necessarily guarantee predictive consistency across all observables. The CPL parametrization appears to achieve its improved $\chi^2$ by fitting certain regions of the data at the expense of others, whereas the QDEE model provides a more balanced description across both global and local consistency tests. This behavior suggests that the truncated Taylor expansion of the DE sector offers a more stable phenomenological framework for describing the late-time expansion history within the current observational precision.

\section{Conclusion}
\label{conclusion}

In this work, we have combined early- and late-time cosmological datasets, including CMB, DESI DR2, and PP supernova measurements to assess the observational viability of the QDEE framework compared to $\Lambda$CDM and CPL as a late-time extension of standard cosmology. Our results show that the fundamental early-Universe parameters remain remarkably stable across all model variants, confirming that the additional DE degrees of freedom introduced in QDEE primarily modify the late-time expansion history while preserving the well-tested physics of recombination and the CMB acoustic scale.

Although the CPL parametrization often achieves the lowest minimum $\Delta \chi^2$ values, this apparent improvement is accompanied by substantial parameter degeneracies and physically unstable behavior. In particular, when DESI DR2 data are included, CPL systematically drives the inferred Hubble constant toward significantly lower values, thereby exacerbating the discrepancy with local distance-ladder measurements. By contrast, the QDEE model produces a more moderate upward shift in $H_0$, partially alleviating Hubble tension relative to $\Lambda\mathrm{CDM}$ while maintaining physically reasonable matter densities and avoiding extreme DE evolution. This behavior suggests that QDEE offers a more controlled and observationally consistent late-time modification.

Bayesian evidence calculations further strengthen this conclusion. Across multiple dataset combinations, QDEE is statistically preferred over standard $\Lambda\mathrm{CDM}$, with evidence differences corresponding to strong support under Jeffreys’ scale. This indicates that the truncated Taylor expansion of the DE density provides an efficient improvement in fit without introducing excessive parameter-space volume or unnecessary model complexity. While CPL can improve raw likelihoods, its broader flexibility is penalized by Bayesian model selection, reflecting reduced predictive efficiency.

Posterior predictive consistency tests provide an additional layer of model validation. Although CPL performs adequately under the global $\chi^2$ PPC statistic, it fails more stringent maximum-discrepancy tests, revealing that its improved fit is not uniformly distributed across datasets and that some observational subsets are poorly described. In contrast, both $\Lambda\mathrm{CDM}$ and QDEE remain broadly consistent with posterior predictive expectations, with QDEE showing only mild residual tension. This demonstrates that QDEE achieves improved flexibility while preserving overall predictive coherence. We also point out that the posterior predictive p-values extremely close to unity should not automatically be interpreted as evidence of superior model performance. Such values may also reflect reduced residual variance relative to the replicated realizations or sensitivity to the adopted covariance structure and test statistic. Consequently, the PPC analysis presented here should be regarded as a complementary consistency diagnostic rather than a definitive model-selection criterion. In this context, the PPC results mainly indicate that the QDEE framework remains statistically compatible with the observed datasets while avoiding the stronger localized discrepancies observed in the CPL parametrization.

Taken together, these findings suggest that while simple dynamical DE parametrizations such as CPL can offer increased phenomenological freedom, they may also produce unstable, dataset-sensitive, or physically problematic cosmological inferences. The QDEE framework instead emerges as a more robust phenomenological extension: it preserves early-Universe consistency, allows modest and controlled departures from a pure cosmological constant, partially relieves late-time cosmological tensions, and is statistically favored by current data.

Future cosmological surveys with improved precision at intermediate redshifts will be crucial for further constraining higher-order DE contributions and determining whether physically motivated density-based parameterizations such as QDEE can provide a genuinely superior description of cosmic expansion beyond the standard $\Lambda\mathrm{CDM}$ paradigm.

\begin{acknowledgments}
S.K. (the first author)\ acknowledges the financial support and access to computational resources provided by Plaksha University. A.J.S.C.\ is partially supported by the Conselho Nacional de Desenvolvimento Cient\'{\i}fico e Tecnol\'ogico (CNPq; National Council for Scientific and Technological Development) under Grant No.~305881/2022-1, and by the Funda\c{c}\~ao da Universidade Federal do Paran\'a (FUNPAR; Paran\'a Federal University Foundation) through public notice 04/2023-Pesquisa/PRPPG/UFPR (Process No.~23075.019406/2023-92). Additional support is provided by the NAPI ``Fen\^omenos Extremos do Universo'' program of the Funda\c{c}\~ao de Apoio \`a Ci\^encia, Tecnologia e Inova\c{c}\~ao do Paran\'a (NAPI F\'ISICA--FASE~2), under protocol No.~22.687.035-0. S.K. acknowledges support from the Startup Research Grant from Plaksha University (File No.\ OOR/PU-SRG/2023-24/08). This work also benefited from the COST Action CA21136 ``Addressing observational tensions in cosmology with systematics and fundamental physics'' (CosmoVerse), supported by COST (European Cooperation in Science and Technology).
\end{acknowledgments}

\bibliography{cite}
\end{document}